\documentclass[12pt]{article}
\usepackage{amsmath,amssymb,amsthm}
\usepackage{graphicx,subfigure}
\usepackage[small]{caption}
\usepackage{changepage}
\usepackage{hyperref,cite,url,breakurl}
\usepackage{mathtools,xparse,MnSymbol}
\usepackage{multirow}
\begin{document}
\baselineskip 0.6cm

\def\bra#1{\langle #1 |}
\def\ket#1{| #1 \rangle}
\def\inner#1#2{\langle #1 | #2 \rangle}
\def\brac#1{\llangle #1 \|}
\def\ketc#1{\| #1 \rrangle}
\def\innerc#1#2{\llangle #1 \| #2 \rrangle}
\def\app#1#2{%
  \mathrel{%
    \setbox0=\hbox{$#1\sim$}%
    \setbox2=\hbox{%
      \rlap{\hbox{$#1\propto$}}%
      \lower1.1\ht0\box0%
    }%
    \raise0.25\ht2\box2%
  }%
}
\def\approxprop{\mathpalette\app\relax}
\DeclarePairedDelimiter{\norm}{\lVert}{\rVert}

\begin{titlepage}

\begin{flushright}
\end{flushright}

\vskip 1.2cm

\begin{center}
{\Large \bf Spacetime and Universal Soft Modes\\
\vspace{1mm}
--- Black Holes and Beyond}

\vskip 0.7cm

{\large Yasunori Nomura}

\vskip 0.5cm

{\it Berkeley Center for Theoretical Physics, Department of Physics,\\
  University of California, Berkeley, CA 94720, USA}

\vskip 0.2cm

{\it Theoretical Physics Group, Lawrence Berkeley National Laboratory, 
 Berkeley, CA 94720, USA}

\vskip 0.2cm

{\it Kavli Institute for the Physics and Mathematics of the Universe 
 (WPI),\\
 UTIAS, The University of Tokyo, Kashiwa, Chiba 277-8583, Japan}

\vskip 0.8cm

\abstract{Recently, a coherent picture of the quantum mechanics of an 
 evaporating black hole has been presented which reconciles unitarity 
 with the predictions of the equivalence principle.  The thermal nature 
 of a black hole as viewed in a distant reference frame arises from 
 entanglement between the hard and soft modes, generated by the chaotic 
 dynamics at the string scale.  In this paper, we elaborate on this 
 picture, particularly emphasizing the importance of the chaotic nature 
 of the string (UV) dynamics across all low energy species in generating 
 large (IR) spacetime behind the horizon.  Implications of this UV/IR 
 relation include $O(1)$ breaking of global symmetries at the string 
 scale and a self-repair mechanism of black holes restoring the smoothness 
 of their horizons.  We also generalize the framework to other systems, 
 including Rindler, de~Sitter, and asymptotically flat spacetimes, and 
 find a consistent picture in each case.  Finally, we discuss the origin 
 of the particular construction adopted in describing the black hole 
 interior as well as the outside of a de~Sitter horizon.  We argue that 
 the construction is selected by the quantum-to-classical transition, 
 in particular the applicability of the Born rule in a quantum mechanical 
 world.}

\end{center}
\end{titlepage}

\tableofcontents
\vspace{1cm}

\section{Introduction}
\label{sec:intro}

Ever since the thermodynamics of a black hole was 
discovered~\cite{Bekenstein:1973ur,Hawking:1974sw}, it has been 
a key element in advancing our understanding of quantum gravity. 
On one hand, the fact that the entropy of a black hole is proportional 
to its horizon area has led to the idea of holography~\cite{'tHooft:1993gx,%
Susskind:1994vu,Bousso:1999xy}, which is elegantly realized in the 
AdS/CFT correspondence~\cite{Maldacena:1997re}.  On the other hand, 
the fact that a black hole has a nonzero temperature has led to 
confusion about the consistency between quantum mechanics and general 
relativity~\cite{Hawking:1976ra,Mathur:2009hf,Almheiri:2012rt}.  It 
is widely believed that a solution to this problem would give us major 
insight into how quantum gravity works at the fundamental level.

In recent work, we have presented a coherent picture of the quantum 
mechanics of an evaporating black hole~\cite{Nomura:2018kia}, building 
on the tools and ideas developed earlier~\cite{Papadodimas:2012aq,%
Maldacena:2013xja,Nomura:2014woa} (for more complete references, see 
Ref.~\cite{Nomura:2018kia}).  From the point of view of a distant 
observer, the thermality of a black hole arises because a vast majority 
of degrees of freedom---which we call soft modes---become temporarily 
unobservable because of the large redshift.  In an ordinary statistical 
mechanical system, an equilibrated state is described by a thermal density 
matrix if we focus only on a small subset, even if the entire state is 
pure.  The same occurs for a black hole.  Since the degrees of freedom 
described by semiclassical theory---which are the complement of the soft 
modes and called hard modes---are only a small subset of the whole, their 
state is given by a thermal density matrix if the black hole is in its 
ground state.  Unlike many other applications in statistical mechanics, 
however, a useful separation between the relevant subsystem and the 
rest in the black hole case is given in momentum space, rather than 
in position space.  The thermality of the hard modes then arises from 
the entanglement in momentum space, i.e.\ between the hard and soft 
modes, resulting from the energy constraint.

The purpose of this paper is threefold.  First, we elucidate the picture 
of Ref.~\cite{Nomura:2018kia}, particularly emphasizing the role played by 
the chaotic dynamics at the string scale.  We find an intriguing relation 
between the chaotic dynamics in the ultraviolet (UV) and the emergence 
of smooth spacetime in the infrared (IR).  This relation suggests that 
the string scale dynamics is chaotic across all low energy species and 
breaks all (linearly realized) global symmetries with $O(1)$ strength 
at the string scale.  We also discuss why the present framework does not 
suffer from the ``Born rule problem'' of Ref.~\cite{Marolf:2015dia} and 
describe a mechanism with which a black hole ``self-repairs'' itself to 
restore the smooth horizon.

We then generalize the framework to other spacetimes, including Rindler, 
de~Sitter, and asymptotically flat spacetimes.  We obtain a consistent 
picture in each case.  In particular, in de~Sitter spacetime, a construction 
similar to the black hole interior allows for describing the situation 
in which semiclassical information going outside the de~Sitter horizon 
is retrieved later when the system tunnels out of the original de~Sitter 
vacuum.  Soft modes play an important role in this description.  In the 
flat space limit, these modes decouple from the dynamics occurring in 
a finite spacetime region.  They, however, seem to be related to the 
existence of an infinite-dimensional asymptotic symmetry group, including 
the (diagonal) Bondi-Metzner-Sachs (BMS) group~\cite{Bondi:1962px,%
Sachs:1962wk,Strominger:2013jfa}.

Finally, we discuss the origin of the particular construction adopted in 
describing the black hole interior and the outside of a de~Sitter horizon. 
We argue that this construction is selected by the quantum-to-classical 
transition, specifically by the requirement of making most manifest the 
observables to which the Born rule can be applied.  While this issue is 
irrelevant for asymptotically flat or AdS spacetime, it can become very 
important when describing a system that is (effectively) finite dimensional, 
such as the black hole interior and cosmological spacetimes.

The organization of this paper is as follows.  In Section~\ref{sec:pic}, 
we describe the framework of Ref.~\cite{Nomura:2018kia}, emphasizing 
its salient feature of the UV/IR relation.  This section forms the basis 
for the rest of the paper.  In Section~\ref{sec:beyond}, we generalize 
the framework to other spacetimes, including Rindler, de~Sitter, and 
asymptotically flat spacetimes.  In Section~\ref{sec:discuss}, we discuss 
the issue of observables in a quantum theory, where the emergence of the 
construction relevant for the black hole interior is elucidated from the 
viewpoint of the applicability of the Born rule.

The framework discussed in this paper is consistent with and, in fact, 
complementary to recent analyses of an evaporating black hole in the 
AdS/CFT correspondence~\cite{Penington:2019npb,Almheiri:2019psf}.%
\footnote{For important differences between the case of a flat 
 space black hole discussed in this paper and the case of 
 Refs.~\cite{Penington:2019npb,Almheiri:2019psf} corresponding 
 to a large AdS black hole, see Ref.~\cite{Nomura:2019dlz}.}
Our analysis elucidates the very appearance of the interior region---in 
particular how the effective second exterior degrees of freedom emerge---from 
a microscopic point of view.  It illuminates what each element in the 
AdS/CFT-based models, involving a large AdS black hole coupled to an 
external auxiliary system, corresponds to in a more realistic black hole. 
The general analysis presented here also sheds light on aspects of 
spacetime in quantum gravity in broader contexts.

Throughout the paper, $l_{\rm P}$ denotes the Planck length. 
We adopt natural units $c = \hbar = 1$.

\section{Black Hole Interior and Universal Soft Modes}
\label{sec:pic}

In this section, we describe the framework of Ref.~\cite{Nomura:2018kia}, 
elucidating several points that play central roles in our discussion. 
We particularly emphasize the importance of the chaotic nature of string 
dynamics across all low energy species and the intriguing relation 
between the dynamics in the UV (string scale) and the emergence of 
smooth spacetime in the IR.  We also explain why the Born rule problem 
of Ref.~\cite{Marolf:2015dia} does not apply to the current framework.

We will focus on Schwarzschild black holes in 4-dimensional asymptotically 
flat spacetime (or small black holes in 4-dimensional asymptotically AdS 
spacetime), though the restriction on specific spacetime dimensions or 
on non-rotating, non-charged black holes is not essential.

\subsection{Evaporating black hole and its interior}
\label{subsec:BH}

We begin our discussion with the description of a black hole as viewed from 
a distant reference frame.  This corresponds to a semiclassical description 
that uses a metric covering the exterior of the black hole, i.e.\ the one 
using the Schwarzschild time (or time analogous to it).  This description arises 
naturally in holography---it is the description obtained from the boundary 
picture through simple bulk reconstruction of renormalization-group/tensor-network 
type~\cite{Nomura:2018kji} and, as such, is unitary.  On the other hand, the 
description suitable for an infalling observer, corresponding to the Kruskal 
extension, is obtained only effectively, as we will see later.

A key observation of Ref.~\cite{Nomura:2018kia} is that the thermal nature 
of a black hole in a distant description arises from entanglement between 
hard and soft modes of low energy quantum fields.%
\footnote{Here and below, low energy fields mean quantum fields existing 
 below the string scale, $1/l_{\rm s}$.}
A mode of a low energy quantum field in the zone region (also called the 
thermal atmosphere)
\begin{equation}
  r_{\rm s} \leq r \leq r_{\rm z}
\label{eq:zone}
\end{equation}
is classified as a hard or soft mode, depending on whether its frequency 
$\omega$, as measured in the asymptotic region, is larger or smaller than
\begin{equation}
  \Delta \approx O\biggl(\frac{1}{M l_{\rm P}^2}\biggr).
\label{eq:Delta}
\end{equation}
Here, $r_{\rm z} \approx 3 M l_{\rm P}^2$, and $r_{\rm s}$ is the location 
of the stretched horizon~\cite{Susskind:1993if} given by%
\footnote{In this paper, we use the $\sim$ symbol to mean equality up to 
 numerical coefficients.}
\begin{equation}
  r_{\rm s} - 2 M l_{\rm P}^2 
  \sim \frac{l_{\rm s}^2}{M l_{\rm P}^2}.
\label{eq:stretched}
\end{equation}
In a distant description, the classical spacetime picture is applicable 
only outside the stretched horizon, and its location is determined by 
the condition that the proper distance from the mathematical horizon, 
$r = 2 M l_{\rm P}^2$, is of order the string length, $l_{\rm s}$.

The quantity $\Delta$ in Eq.~(\ref{eq:Delta}) is taken to be somewhat, 
e.g.\ by a factor of $O(10)$, larger than the Hawking temperature
\begin{equation}
  T_{\rm H} = \frac{1}{8\pi M l_{\rm P}^2}.
\label{eq:T_H}
\end{equation}
The separation of the modes described above is motivated by the fact that 
the configuration of soft modes cannot be determined operationally by 
a physical probe within the characteristic timescale with which the 
internal state of the system varies.  Note that $\Delta$ is the inverse 
of the timescale for single Hawking emission, so that the uncertainty 
principle prevents us from specifying the energy of the black hole state 
better than that.  Below, we will assume that the energy (mass) of a black 
hole is specified with this maximal precision.

A state of a black hole---representing the state of the system in the 
black hole zone region---is given in its semiclassical vacuum by a generic 
state consistent with the energy constraint imposed on the region.  In 
particular, in an idealized limit in which the black hole is isolated 
from the environment, a state of a black hole of mass $M$ is given by
\begin{equation}
  \ket{\Psi(M)} = \sum_n \sum_{i_n = 1}^{{\cal N}(M-E_n)} 
    c_{n i_n} \ket{\{ n_\alpha \}} \ket{\psi_{i_n}(M-E_n)},
\label{eq:BH-state}
\end{equation}
where $\ket{\{ n_\alpha \}}$ are orthonormal states of the hard modes 
with $n \equiv \{ n_\alpha \}$ representing the set of all occupation 
numbers; $\alpha$ collectively denotes the species, frequency, and 
angular-momentum quantum numbers of a mode, and $E_n$ is the energy of 
the state $\ket{\{ n_\alpha \}}$ as measured in the asymptotic region 
(with precision $\Delta$).  $\ket{\psi_{i_n}(M-E_n)}$ are orthonormal 
states of the soft modes which have energy $M-E_n$, with the index $i_n$ 
running from $1$ to ${\cal N}(M-E_n)$, where
\begin{equation}
  {\cal N}(M) = e^{S_{\rm BH}(M)} \frac{\Delta}{M} 
  = e^{4\pi M^2 l_{\rm P}^2} \frac{\Delta}{M}.
\label{eq:cal-N}
\end{equation}
Below, we ignore the logarithmic correction of order $\ln(\Delta/M)$ 
to the entropy, identifying ${\cal N}(M)$ as the density of states 
$e^{S_{\rm BH}(M)}$.  We also assume that all the states are normalized; 
for example, we assume $\sum_n \sum_{i_n = 1}^{{\cal N}(M-E_n)} 
|c_{n i_n}|^2 = 1$ in Eq.~(\ref{eq:BH-state}).  Note that the total 
entropy of the state of the form in Eq.~(\ref{eq:BH-state}) is
\begin{equation}
  \ln\biggl[ \sum_n e^{S_{\rm BH}(M-E_n)} \biggr] \approx S_{\rm BH}(M),
\label{eq:S-total}
\end{equation}
so that the logarithm of the number of independent black hole microstates 
is given by the standard Bekenstein-Hawking entropy.

In a realistic setup, a black hole state is entangled with the environment 
which generally involves Hawking radiation emitted earlier.  The state of 
the total system is then given by
\begin{equation}
  \ket{\Psi(M)} = \sum_n \sum_{i_n = 1}^{{\cal N}(M-E_n)} 
    \sum_a c_{n i_n a} \ket{\{ n_\alpha \}} 
    \ket{\psi_{i_n}(M-E_n)} \ket{\phi_a},
\label{eq:sys-state}
\end{equation}
where $\ket{\phi_a}$ represents the state of the system in the far region 
$r > r_{\rm z}$.  Note that states of this form appear in a distant 
description of the system, which corresponds to the boundary description 
in a holographic theory, so that their evolution is unitary.  The thermal 
nature of a black hole arises because, within the zone, semiclassical 
theory describes microscopic dynamics of only the hard modes.  Indeed, 
tracing out soft modes in Eq.~(\ref{eq:sys-state}) yields
\begin{equation}
  {\rm Tr}_{\rm soft} \ket{\Psi(M)} \bra{\Psi(M)}
  = \frac{1}{\sum_n e^{-\frac{E_n}{T_{\rm H}}}} \sum_n 
    e^{-\frac{E_n}{T_{\rm H}}} \ket{\{ n_\alpha \}} \bra{\{ n_\alpha \}} 
    \otimes \rho_{\phi,n},
\label{eq:rho_HF}
\end{equation}
where we have assumed that the coefficients $c_{n i_n a}$ take generic 
values in the spaces of the hard and soft modes; i.e., black hole states 
are generic.  $\rho_{\phi,n}$ are ($n$-dependent) reduced density matrices 
for the far modes.

A small object in the zone region, with the characteristic size $d$ in the 
angular directions much smaller than the horizon, $d \ll M l_{\rm P}^2$, 
can be described by annihilation and creation operators acting only on 
the hard modes~\cite{Nomura:2018kia}:
\begin{align}
  b_\gamma &= \sum_n \sqrt{n_\gamma}\, 
    \ket{\{ n_\alpha - \delta_{\alpha\gamma} \}} \bra{\{ n_\alpha \}},
\label{eq:ann}\\
  b_\gamma^\dagger &= \sum_n \sqrt{n_\gamma + 1}\, 
    \ket{\{ n_\alpha + \delta_{\alpha\gamma} \}} \bra{\{ n_\alpha \}}.
\label{eq:cre}
\end{align}
What is the fate of such an object falling toward the black hole?

In a distant/boundary description, in which the evolution of a state can 
be unitary for arbitrarily long time, a small object falling into the 
black hole first becomes excitations of the stretched horizon, whose 
information will then be dissipated into the state of the soft modes 
and eventually sent back to ambient space by Hawking emission.  This 
description, however, is not suitable for finding what the object falling 
into the black hole will actually see.  Because of a large relative 
boost between the object and the distant frame, which formally becomes 
infinite as the object approaches the horizon, macroscopic time 
experienced by the object is mapped to an extremely short time when 
measured by a stationary observer at the location of the object.  This 
implies that the experience of the object occurs ``instantaneously'' 
in a distant description (of order the cutoff time for an observer at 
$r = r_{\rm s}$).  Understanding it, therefore, requires time evolution 
different from the boundary one, specifically an evolution associated 
with the proper time of the object.

Such a picture---an infalling description---is obtained after coarse-graining 
the soft (and associated far) mode degrees of freedom~\cite{Nomura:2018kia,%
Papadodimas:2012aq,Maldacena:2013xja}, which cannot be discriminated by 
a fallen object in a timescale available to it.  Suppose that at a boundary 
time $t_*$, the state of the system is given by Eq.~(\ref{eq:sys-state}) 
with $c_{n i_n a}$ taking generic values in the $n$ and $i_n$ spaces.%
\footnote{In an asymptotically flat spacetime, the boundary time $t$ can be 
 taken as the Schwarzschild time.  In more general cases, $t$ can be a time 
 parameter on the holographic screen~\cite{Bousso:1999cb,Nomura:2016ikr}.}
We can then erect an effective theory {\it based on this state} by 
coarse-graining the soft and far modes:
\begin{equation}
  \sum_{i_n = 1}^{{\cal N}(M-E_n)} \sum_a c_{n i_n a} 
    \ket{\psi_{i_n}(M-E_n)} \ket{\phi_a} 
  \longrightarrow\
  \frac{e^{-\frac{E_n}{2 T_{\rm H}}}}
      {\sqrt{\sum_n e^{-\frac{E_n}{T_{\rm H}}}}}\, \ketc{\{ n_\alpha \}},
\label{eq:coarse}
\end{equation}
where we have used the same label as the corresponding hard mode state to 
specify the coarse-grained state, which we denote by the double ket symbol, 
and the coefficient in the right-hand side arises from the normalization 
condition for $\ketc{\{ n_\alpha \}}$.  The state in Eq.~(\ref{eq:sys-state}) 
in this effective theory is then given by
\begin{equation}
  \ketc{\Psi(M)} 
  = \frac{1}{\sqrt{\sum_n e^{-\frac{E_n}{T_{\rm H}}}}} 
    \sum_n e^{-\frac{E_n}{2 T_{\rm H}}} 
    \ket{\{ n_\alpha \}} \ketc{\{ n_\alpha \}},
\label{eq:BH-coarse}
\end{equation}
which takes the form of the standard thermofield double state in the 
two-sided black hole picture~\cite{Unruh:1976db,Israel:1976ur}, although 
$\ket{\{ n_\alpha \}}$ here represent the states only of the hard modes.

We emphasize that in order to obtain the correct Boltzmann-weight 
coefficients in Eqs.~(\ref{eq:coarse},~\ref{eq:BH-coarse}), $\propto 
e^{-E_n/2 T_{\rm H}}$, it is important that the black hole has soft 
modes with the density of states given by $e^{S_{\rm BH}(E_{\rm soft})}$, 
and that the hard and soft modes are well scrambled, giving $c_{n i_n a}$ 
that take values statistically independent of $n$.%
\footnote{Note that even after the Page time~\cite{Page:1993wv}, when 
 the {\it coarse-grained} entropy of the emitted radiation is greater 
 than the black hole entropy, the number of independent microstates that 
 couple to the hard mode state $\ket{\{ n_\alpha \}}$ is still controlled 
 by the density of soft mode states, $e^{S_{\rm BH}(M-E_n)}$.}
We also stress that the operation of coarse-graining, i.e.\ {\it ignoring} 
the detailed structure of $c_{n i_n a}$'s, is different from tracing 
out degrees of freedom.  It is this coarse-graining that leads to 
the apparent uniqueness of the infalling vacuum, despite the existence 
of exponentially many black hole microstates.

We can now define the ``mirror operators'' acting on the coarse-grained 
states
\begin{align}
  \tilde{b}_\gamma &= \sum_n \sqrt{n_\gamma}\, 
    \ketc{\{ n_\alpha - \delta_{\alpha\gamma} \}} 
    \brac{\{ n_\alpha \}},
\label{eq:ann-m}\\
  \tilde{b}_\gamma^\dagger &= \sum_n \sqrt{n_\gamma + 1}\, 
    \ketc{\{ n_\alpha + \delta_{\alpha\gamma} \}} 
    \brac{\{ n_\alpha \}}.
\label{eq:cre-m}
\end{align}
This implies that modes in the second exterior of the effective theory 
arise as (hard) quasi-particles, generated by collective excitations 
of the soft modes as well as the far mode degrees of freedom entangling 
with them, including early Hawking radiation.  Note that at 
the microscopic level, these operators act on both soft 
and far degrees of freedom.  Indeed, tracing out the soft/far modes 
in Eq.~(\ref{eq:sys-state}), the remaining correlation between the 
hard and far/soft modes is essentially classical.

The mirror operators in Eqs.~(\ref{eq:ann-m},~\ref{eq:cre-m}) allow us, 
together with the operators in Eqs.~(\ref{eq:ann},~\ref{eq:cre}), to form 
the annihilation and creation operators for infalling modes:
\begin{align}
  a_\xi &= \sum_\gamma 
    \bigl( \alpha_{\xi\gamma} b_\gamma 
    + \beta_{\xi\gamma} b_\gamma^\dagger 
    + \zeta_{\xi\gamma} \tilde{b}_\gamma 
    + \eta_{\xi\gamma} \tilde{b}_\gamma^\dagger \bigr),
\label{eq:a_xi}\\
  a_\xi^\dagger &= \sum_\gamma 
    \bigl( \beta_{\xi\gamma}^* b_\gamma 
    + \alpha_{\xi\gamma}^* b_\gamma^\dagger 
    + \eta_{\xi\gamma}^* \tilde{b}_\gamma 
    + \zeta_{\xi\gamma}^* \tilde{b}_\gamma^\dagger \bigr),
\label{eq:a_xi-dag}
\end{align}
where $\xi$ is the label in which the frequency $\omega$ with respect to 
$t$ is traded with the frequency $\Omega$ associated with the infalling 
time, and $\alpha_{\xi\gamma}$, $\beta_{\xi\gamma}$, $\zeta_{\xi\gamma}$, 
and $\eta_{\xi\gamma}$ are the Bogoliubov coefficients calculable using 
the standard field theory method.  The generator of time evolution in 
this description is then given by
\begin{equation}
  H = \sum_\xi \Omega a_\xi^\dagger a_\xi 
    + H_{\rm int}\bigl( a_\xi, a_\xi^\dagger \bigr).
\label{eq:H_eff}
\end{equation}
This leads to the physics of a smooth horizon.  The existence of these 
operators implies that there is a subsector in the original microscopic 
theory encoding the experience of an object after it crosses the horizon. 
(For further discussion, see Section~\ref{sec:discuss}.)

The effective theory erected as above is applicable only for a limited 
spacetime region~\cite{Nomura:2018kia}.  Since the far modes are 
coarse-grained, the theory describes only physics within the causal 
domain of the union of the zone and its mirror regions on the $t = t_*$ 
hypersurface (the time at which the effective theory is erected). 
Furthermore, the fact that the soft modes are coarse-grained implies 
that the description is intrinsically semiclassical; i.e., it is valid 
only down to the lengthscale $l_{\rm s}$.  This suggests that the 
singularity of a black hole may not be resolved; it may simply 
represent an intrinsic limitation coming from the fact that the theory 
of the interior is obtained by coarse-graining and hence describes 
a finite-dimensional, non-unitary system.  The fact that an effective 
theory describes only a limited spacetime region also implies that the 
picture of the whole interior, as indicated by general relativity, can 
be obtained only by using multiple effective theories erected at different 
times (which are generally not independent).  This is the sense in which 
the concept of the black hole interior emerges from the microscopic 
description of the black hole.

\subsection{``Spacetime'' and matter within low energy fields}
\label{subsec:spacetime-matter}

A salient feature of the framework described above is that low energy 
quantum fields contain both degrees of freedom associated with the 
Bekenstein-Hawking entropy (spacetime) and matter (excitations).  One 
might be skeptical about this, but there are many arguments suggesting 
that it indeed gives a valid picture.

\subsubsection*{The number of low energy species}

Let us estimate the number of soft modes contained in a single low energy 
field.  This is done by integrating the entropy density
\begin{equation}
  s_0(r) = c\, T_{\rm loc}(r)^3
\label{eq:s-density}
\end{equation}
over the zone region, Eq.~(\ref{eq:zone}), where $c$ is a constant of $O(1)$, 
and 
\begin{equation}
  T_{\rm loc}(r) = \frac{T_{\rm H}}{\sqrt{1-\frac{2M l_{\rm P}^2}{r}}}
\label{eq:T_loc}
\end{equation}
is the local temperature measured at $r$.  This gives
\begin{equation}
  S_{{\rm soft},0} = \int_{r_{\rm s}}^{r_{\rm z}}\! s_0(r)\, 
    \frac{r^2 dr d\Omega}{\sqrt{1-\frac{2M l_{\rm P}^2}{r}}} 
  \sim \frac{M^2 l_{\rm P}^4}{l_{\rm s}^2}.
\label{eq:S_soft,0}
\end{equation}
Alternatively, one could directly count the number of modes excited. 
Specifically, a particle with angular momentum $L^2 = \ell(\ell+1)$ costs 
the energy, as measured in the asymptotic region, of
\begin{equation}
  \varDelta \omega \sim \frac{\ell}{r},
\label{eq:ang-cost}
\end{equation}
so that
\begin{equation}
  \frac{\varDelta \omega}{T_{\rm loc}(r)} 
  \sim \frac{\ell}{T_{\rm H} r} \sqrt{1-\frac{2Ml_{\rm P}^2}{r}} 
  \sim \ell \sqrt{1-\frac{2Ml_{\rm P}^2}{r}}.
\label{eq:ang-TH}
\end{equation}
Therefore, modes up to $\ell_{\rm max} \sim \sqrt{r/(r-2Ml_{\rm P}^2)}$ are 
effectively populated, giving the same result as Eq.~(\ref{eq:S_soft,0}):
\begin{equation}
  S_{{\rm soft},0} 
  \sim \sum_{\ell=0}^{\ell_{\rm max}}(2\ell+1) \Big|_{r = r_{\rm s}} 
  \sim \frac{Ml_{\rm P}^2}{r_{\rm s} - 2Ml_{\rm P}^2} 
  \sim \frac{M^2 l_{\rm P}^4}{l_{\rm s}^2}.
\label{eq:S_soft,0-2}
\end{equation}
By going from the angular momentum to angular position bases, we find 
that most of the soft mode degrees of freedom are located on the stretched 
horizon, with $O(1)$ degrees of freedom per string area $\sim l_{\rm s}^2$ 
for each low energy field.

The total entropy of the soft modes is given by multiplying the number of 
low energy fields, $N$, to $S_{{\rm soft},0}$:
\begin{equation}
  S_{\rm soft} \sim N S_{{\rm soft},0} 
  \sim N \frac{M^2 l_{\rm P}^4}{l_{\rm s}^2}.
\label{eq:S_soft}
\end{equation}
Using the relation expected in any theory of quantum gravity (see, e.g., 
Ref.~\cite{Dvali:2007hz})
\begin{equation}
  l_{\rm P}^2 \sim \frac{l_{\rm s}^2}{N},
\label{eq:qg-rel}
\end{equation}
we find that this indeed reproduces the Bekenstein-Hawking entropy, up to 
an incalculable $O(1)$ factor:
\begin{equation}
  S_{\rm soft} \sim M^2 l_{\rm P}^2 \sim S_{\rm BH}.
\label{eq:S-agree}
\end{equation}
This is consistent with the view that the Bekenstein-Hawking entropy 
is mostly on the stretched horizon with the surface entropy density of 
$1/4 l_{\rm P}^2 \sim N/l_{\rm s}^2$.

The fact that the distribution of the soft modes strongly peaks toward the 
stretched horizon implies the existence of an arbitrariness in splitting 
them into parts in ``high energy'' (i.e.\ horizon) and low energy degrees 
of freedom.  In this paper, we adopt a scheme in which the entire 
Bekenstein-Hawking entropy is associated with the soft modes of the 
low energy fields, which was already implied when we took the density 
of soft mode states to be $e^{S_{\rm BH}(E_{\rm soft})}$.  This is 
consistent because we do not describe the internal dynamics among soft 
modes.  In fact, we know that the dynamics of the majority of the soft 
modes cannot be described by the low energy theory because they are localized 
near/at the stretched horizon, where the local intrinsic scale for the 
dynamics is of order the string scale $1/l_{\rm s}$.  The internal dynamics 
of these modes is indeed expected to be nonlocal in the directions along 
the horizon~\cite{Hayden:2007cs,Sekino:2008he}.%
\footnote{This feature may be used to discriminate the hard modes 
 from soft modes near the stretched horizon in a holographic theory 
 at a boundary.  A similar separation of modes has been discussed 
 in Ref.~\cite{Balasubramanian:2018yjq} for a half-BPS ``superstar'' 
 geometry in AdS/CFT.}

Note the crucial role played by the fact that the Bekenstein-Hawking 
entropy is distributed universally over all the low energy fields. 
It is this feature that reconciles the fact that the spacetime picture 
breaks down at the string scale (at $r \sim r_{\rm s}$ where $T_{\rm loc} 
\sim 1/l_{\rm s}$) with the fact that $S_{\rm BH}$ can be written in terms 
of the Planck length, $l_{\rm P}$, without involving $l_{\rm s}$.

\subsubsection*{Bekenstein bound}

An object in field theory in the near horizon region obeys the bound, 
first envisioned by Bekenstein~\cite{Bekenstein:1980jp} and proved in 
Ref.~\cite{Casini:2008cr}:
\begin{equation}
  S \lesssim 2\pi E_{\rm loc} \rho,
\label{eq:B-bound}
\end{equation}
where $S$ and $E_{\rm loc}$ are the entropy and energy of the object (with 
$E_{\rm loc}$ being measured locally at the location of the object), and 
$\rho$ is the proper distance between the object and the horizon.  We expect 
that an excitation above the field theory vacuum has
\begin{equation}
  S \gtrsim c,
\label{eq:S-exc}
\end{equation}
where $c$ is a number of order a few,%
\footnote{Precisely speaking, $S$ in Eq.~(\ref{eq:B-bound}) represents 
 the difference of the von~Neumann entropy between the excited and vacuum 
 states.  In the analysis below, we coarse-grain the excitation sufficiently 
 so that the resulting mixed state has $S$ sufficiently larger than $1$.}
and thus
\begin{equation}
  c \,\lesssim\, S 
  \,\lesssim\, 2\pi E_{\rm loc} \rho 
  \,\approx\, \frac{E_{\rm loc}}{T_{\rm loc}}.
\label{eq:ineq}
\end{equation}
Here, $T_{\rm loc}$ is the local Hawking temperature at the location of 
the object, and we have used the fact that $\rho \approx 1/2\pi T_{\rm loc}$. 
We find that an excitation in field theory must have energy
\begin{equation}
  E \,\gtrsim\, c T_{\rm H} 
  \,\approx\, O\biggl(\frac{1}{M l_{\rm P}^2}\biggr),
\label{eq:hard-cond}
\end{equation}
as measured in the asymptotic region.  This is exactly the condition of 
being a hard mode.

An alternative way of viewing this is that Eq.~(\ref{eq:B-bound}) says 
that there are no multiple independent states at the field theory level 
which correspond to the exponentially many states obtained by exciting 
soft modes with $E \ll 1/M l_{\rm P}^2$.  This implies that the 
microstates corresponding to different soft mode excitations must all 
be viewed as the same vacuum state at the level of (semiclassical) 
field theory.

\subsubsection*{Horizon duality:\ chaotic UV dynamics leads to smooth 
IR spacetime}

The appearance of interior spacetime through Eq.~(\ref{eq:coarse}) 
requires a generic black hole state; i.e., the coefficients $c_{n i_n a}$ 
in Eq.~(\ref{eq:sys-state}) take generic values in the $n$ and $i_n$ 
spaces.  This implies that the dynamics of the black hole must be chaotic 
{\it across all low energy species}.  Since the intrinsic dynamics of 
a black hole occurs mostly at the stretched horizon, where the local 
temperature becomes the string scale, this is translated into the 
statement about the string dynamics.  In particular, the dynamics 
at the string scale must not have a structure which prevents the universal 
redistribution of the initial state energy and information over all low 
energy species, such as an exact global symmetry.  In fact, the breaking 
of a global symmetry must be strong, parametrically of $O(1)$, at the 
string scale, in order for the interior spacetime to develop within a 
reasonable time after the black hole formation (or for the black hole 
to self-repair sufficiently quickly; see Section~\ref{subsec:flow}).%
\footnote{I have recently learned that a similar claim about the strength 
 of global symmetry breaking is being pursued by Cordova, Ohmori, and 
 Rudelius using a different, swampland related argument.}
This is consistent with earlier observations, e.g.\ in 
Refs.~\cite{Maldacena:2015waa,Harlow:2018tng}, though it makes 
a stronger statement about the dynamics at the string scale.

Incidentally, a global symmetry that is nonlinearly realized at the string 
scale is not constrained by the argument given above.  In other words, the 
required global symmetry breaking at $l_{\rm s}$ need not be explicit and 
can be ``spontaneous.''  This suggests that the QCD axion needed to solve the 
strong $CP$ problem is a string axion (see, e.g., Ref.~\cite{Svrcek:2006yi}), 
since the required quality of the Peccei-Quinn symmetry is very high. 
If the Peccei-Quinn symmetry were linearly realized at $l_{\rm s}$, 
then the above argument would say that it must be explicitly broken 
with $O(1)$ strength (unless it arises as an approximate accidental 
symmetry at low energies, resulting from a judicious choice of matter 
representations under gauge symmetry), which would invalidate the Peccei-Quinn 
mechanism~\cite{Kamionkowski:1992mf,Barr:1992qq,Holman:1992us}.

It is important that the universal dynamics discussed above, leading to 
generic entanglement between the hard and soft modes, emerges only if the 
surface of the material composing an object recedes behind the surface 
at which $T_{\rm loc} \sim 1/l_{\rm s}$, i.e.\ the stretched horizon. 
This condition, therefore, can be used as a criterion for differentiating 
a black hole from normal matter, such as a piece of coal and a regular 
star.  For the latter, the structure of the state does not take the 
universal form, even though some radiation may be emitted from its 
surface.  The construction of Eq.~(\ref{eq:coarse}), therefore, does 
not apply, and hence no near-empty interior spacetime.

The analysis here reveals an intriguing relation between IR and UV 
physics:\ in order to have large---IR---spacetime behind a horizon, 
its dynamics as viewed from a distance---the UV dynamics---must be 
chaotic across all low-energy species.  For a quasi-static system, this 
can be stated more quantitatively.  Suppose that a spacetime is given 
with a quasi-static time foliation.  For an evolving black hole, this 
could be done by pulling in ``leaves'' (equal time hypersurfaces on the 
boundary) along holographic slices~\cite{Nomura:2018kji}.  This pulling-in 
procedure must halt when gravity/acceleration becomes large, specifically 
at a surface on which
\begin{equation}
  a \equiv \sqrt{g_{\mu\nu} a^\mu a^\nu} \sim \frac{1}{l_{\rm s}},
\label{eq:acc-mag}
\end{equation}
where
\begin{equation}
  a^\mu = n^\nu \nabla_\nu n^\mu,
\label{eq:acc-def}
\end{equation}
and $n^\mu$ is the timelike unit normal to the holographic---or bulk equal 
time---slice.  Because of large acceleration in Eq.~(\ref{eq:acc-mag}), 
physics on this surface is described by the string scale dynamics. 
It is this dynamics that leads to a well scrambled state in 
Eq.~(\ref{eq:BH-state}), or Eq.~(\ref{eq:sys-state}), allowing 
for the construction of spacetime behind it through Eq.~(\ref{eq:coarse}).

\subsection{The paradox of low energy excitations and its resolution}
\label{subsec:paradox}

In Ref.~\cite{Marolf:2015dia}, an important problem was pointed out 
which was claimed to plague any ``state-dependent'' construction (see 
also Ref.~\cite{Harlow:2014yoa}).  The basic argument is as follows. 
Let us denote the space of pure states with energy $E < E_0$ as 
${\cal H}_{E_0}$.  For sufficiently large $E_0$, we assume that 
a typical state in ${\cal H}_{E_0}$ is a black hole with a smooth 
horizon (which is what the state dependence is supposed to achieve).%
\footnote{Originally, this statement was considered in the context of 
 AdS/CFT duality, where the relevant black hole was a large AdS black 
 hole.  On the other hand, here we are interested in a black hole in 
 asymptotically flat spacetime (or a small AdS black hole).  The analysis 
 below, however, still applies if, instead of CFT Hilbert space, one 
 considers effective Hilbert space ${\cal H}_\Gamma$ on the boundary 
 $\Gamma$ that is pulled in~\cite{Nomura:2018kji} to become a surface 
 near the black hole, e.g.\ the $r = c M l_{\rm P}^2$ surface with 
 $c \gtrsim 3$.  Possible entanglement between the degrees of freedom 
 inside $\Gamma$ and those outside does not play an important role 
 in the discussion here.}
Let us now consider unitaries $U_I$ of the form $e^{i \sum_k \phi_k 
(b_\omega^\dagger b_\omega)^k}$ using appropriately smoothed mode 
operators whose frequencies (as measured in the asymptotic region) are 
$\omega \ll \Delta$.  These operators, because of the smoothing, change 
the energy of the state by $\delta E$, where $|\delta E| \ll \Delta$.

One can then show that a typical state in ${\cal H}_{E_0 + \delta E}$ 
is nearly parallel to a state in $U_I {\cal H}_{E_0}$. 
Reference~\cite{Marolf:2015dia} argues that this is a contradiction. 
The claim is that almost all states in $U_I {\cal H}_{E_0}$ must be 
excited states (stating that otherwise the frozen vacuum argument 
of Ref.~\cite{Bousso:2013ifa} applies), while almost all states in 
${\cal H}_{E_0 + \delta E}$ are vacuum states (as those in ${\cal H}_{E_0}$). 
However, since a state in the latter class is nearly parallel to a state 
in the former class, this leads to a massive violation of the Born rule.

Our framework addresses this issue in a simple (trivial) manner:%
\footnote{For an attempt to address the issue using causality of 
 AdS spacetime, see Ref.~\cite{Raju:2016vsu}.}\
A typical state in $U_I {\cal H}_{E_0}$ is {\it not} an excited state, 
but it is a microstate of the black hole {\it vacuum} which is different 
from those in ${\cal H}_{E_0}$.  Note that given $|\delta E| \ll 
\Delta$, operating $U_I$ on a black hole vacuum state of the form 
in Eq.~(\ref{eq:sys-state}) corresponds to changing the coefficients 
$c_{n i_n a}$.  This simply leads to another black hole vacuum microstate 
having different coefficients $c'_{n i_n a}$.  In other words, one can 
repeat the whole construction of Section~\ref{subsec:BH} with $c_{n i_n a}$ 
replaced with $c'_{n i_n a}$ to find infalling operators $a_\xi^\prime$, 
$a_\xi^{\prime\dagger}$, and $H'$, in terms of which the description 
of a smooth horizon is obtained.  Operations involving structures 
finer than $\Delta$ are not represented by quantum operators in 
semiclassical theory.  (We might say that they correspond to changing 
the background geometry by minuscule amounts, e.g.\ $|\delta M| 
\lesssim \Delta$.)  In the context of holography, soft modes are 
not represented as degrees of freedom specifying states within 
a code subspace~\cite{Almheiri:2014lwa} in a way that subsystem 
recovery is possible.

It is important that exciting a mode with $\omega \gtrsim \Delta$ on 
a black hole vacuum state does not lead to another black hole vacuum 
state; it leads to an excited state.  For example, one can consider 
a state in which there are $N$ particles in the zone:
\begin{equation}
  \ket{\Psi_N} \propto \prod_{i=1}^N
    \left( \sum_\gamma f_{i,\gamma} b_\gamma^\dagger \right) 
    \ket{\Psi(M)},
\label{eq:Psi_N}
\end{equation}
where $f_{i,\gamma}$ are the weights for producing particle $i$ 
by superposing creation operators $b_\gamma^\dagger$ that have 
frequencies larger than $\Delta$.  (Note that for hard particles 
in the zone, $a_\gamma^\dagger \approx b_\gamma^\dagger$.) 
States like Eq.~(\ref{eq:Psi_N}) cannot be written in the form 
of Eq.~(\ref{eq:sys-state}) with the coefficients $c_{n i_n a}$ 
taking generic values in the $n$ and $i_n$ spaces.

In fact, unlike the black hole vacuum states obtained by changing the 
configuration of soft modes, states obtained by exciting hard modes 
are not typical in the Hilbert space.  Consider the space of all 
states that are obtained by acting appropriately smoothed hard mode 
operators on an element of ${\cal H}_{E_0}$ and have energies smaller 
than $E_0 + \omega$, where $\omega \gtrsim \Delta$.  We denote this 
space by $B_\omega {\cal H}_{E_0}$.  One can then show that a typical 
state $\ket{\psi}$ in ${\cal H}_{E_0 + \omega}$ can be written as
\begin{equation}
  \ket{\psi} = \sin\theta\, \ket{\psi_{\rm exc}} 
    + \cos\theta\, \ket{\psi_{\rm vac}},
\label{eq:decomp}
\end{equation}
where $\ket{\psi_{\rm exc}}$ and $\ket{\psi_{\rm vac}}$ are elements of 
$B_\omega {\cal H}_{E_0}$ and its complement ${\cal H}_{E_0 + \omega} 
/ B_\omega {\cal H}_{E_0}$, respectively, and
\begin{equation}
  \sin^2\!\theta \sim e^{-\frac{\omega}{T_{\rm H}}}.
\label{eq:atypical}
\end{equation}
We find that a typical state in ${\cal H}_{E_0 + \omega}$ has only 
negligible overlap with the excited states; i.e., a state obtained 
by exciting hard modes is atypical in the microscopic Hilbert space. 
What the semiclassical theory describes is the dynamics of these 
atypical states.

Another way to phrase the conclusion here is that the 
Hilbert space ${\cal H}_{\rm inf}$ for the infalling modes,  
Eqs.~(\ref{eq:a_xi},~\ref{eq:a_xi-dag}), built on each of the 
black hole and radiation microstates need not overlap with each 
other. As states excited by $b_\gamma^\dagger$'s are atypical 
and can be discriminated from typical states, states excited by 
$a_\xi^\dagger$'s are also atypical in the Hilbert space of the 
black hole and radiation (of which states with thermal Hawking 
radiation are typical) and can be discriminated from typical 
states by energetic consideration.  In particular, as long as 
we are focusing on small excitations, with $\ln\, {\rm dim}\, 
{\cal H}_{\rm inf} \sim ({\cal A}(M)/l_{\rm P}^2)^q$ ($q < 1$) 
where ${\cal A}(M)$ is the area of the black hole horizon, the 
existence of ${\cal H}_{\rm inf}$ is entropically negligible, 
so that it can be attached to each of the black hole vacuum 
microstates without affecting the entropic consideration of black 
hole evaporation at the leading order in $l_{\rm P}^2/{\cal A}(M)$. 
In short, our framework does not employ a type of state dependence 
considered in Ref.~\cite{Marolf:2015dia}.

\subsection{Flow of information and energy: recapitulation}
\label{subsec:flow}

The picture emerging from the analyses described above is the following. 
Gravitational collapse makes a material surface recede until it reaches 
the point at which the local acceleration of a stationary observer 
becomes the string scale, $a \sim 1/l_{\rm s}$.  When this happens, 
chaotic dynamics at the string scale distributes the energy of the 
material into all low energy species.  Indeed, following the earlier 
analysis of the entropy, we can integrate the energy density of the 
soft modes of a single species
\begin{equation}
  \rho_0(r) \sim T_{\rm loc}(r)^4 \sqrt{1-\frac{2Ml_{\rm P}^2}{r}}
\label{eq:soft-rho}
\end{equation}
over the zone region $r_{\rm s} \leq r \leq r_{\rm z}$ (where the second 
factor on the right-hand side is the redshift factor), obtaining
\begin{equation}
  E_{{\rm soft},0} \sim \frac{M l_{\rm P}^2}{l_{\rm s}^2}.
\label{eq:E_soft,0}
\end{equation}
We can therefore reproduce the black hole mass (parametrically) 
after multiplying the number of low energy species $N$.  Furthermore, 
due to the energy constraint, the chaotic dynamics generates 
generic entanglement between the hard and soft modes as in 
Eqs.~(\ref{eq:BH-state},~\ref{eq:sys-state}).  It is this generic 
entanglement that allows for reconstructing spacetime behind the 
horizon through the coarse-graining in Eq.~(\ref{eq:coarse}).

Hawking emission, in the sense of emitting quanta that can be viewed as 
excitations in semiclassical theory, occurs around the edge of the zone:\ 
$|r^*| \lesssim O(M l_{\rm P}^2)$~\cite{Nomura:2014woa}, where
\begin{equation}
  r^* = r + 2M l_{\rm P}^2\, 
    \ln\frac{r-2M l_{\rm P}^2}{2M l_{\rm P}^2}
\label{eq:tortoise}
\end{equation}
is the tortoise coordinate.  An important point is 
that while the distribution of the soft modes is strongly 
peaked toward the stretched horizon, there are $O(1)$ 
degrees of freedom---which are tiny and fractionally only of 
$O(1/M^2 l_{\rm P}^2)$---located around the edge of the zone. 
The information stored in these degrees of freedom is transferred 
to far modes (a Hawking quantum) in this region in each timescale 
of $O(M l_{\rm P}^2)$.  The backreaction creates an ingoing flux 
of negative energy and negative entropy with respect to the static 
(Hartle-Hawking~\cite{Hartle:1976tp}) vacuum.  Note that the only 
relevant low energy fields in this process are those with masses 
smaller than $T_{\rm H}$, since $T_{\rm loc} \sim T_{\rm H}$ around 
the edge of the zone.  The evaporation of the black hole completes 
after $O(M^2 l_{\rm P}^2)$ steps of this elementary emission process. 
The entanglement entropy between the black hole and the radiation 
emitted, $S^{\rm vN}_{\rm hard + soft} = S^{\rm vN}_{\rm rad}$, 
follows the Page curve~\cite{Page:1993wv}, where $S^{\rm vN}_A$ 
represents the von~Neumann entropy of subsystem $A$.

During the evaporation, the structure of entanglement between the hard-mode, 
soft-mode, and radiation components can be written as~\cite{Nomura:2018kia}
\begin{equation}
  \ket{\Psi(M)} = \sum_n \sum_{i_n=1}^{{\cal N}_n} 
    c^n_{i_n} \ket{H_n} \ket{S_{n,i_n}} \ket{R_{n,i_n}},
\label{eq:ent-str}
\end{equation}
where $\ket{H_n}$, $\ket{S_{n,i_n}}$, and $\ket{R_{n,i_n}}$ are states of 
the hard modes, soft modes, and radiation, respectively, and
\begin{equation}
  {\cal N}_n = {\rm min}\{ e^{S_{\rm BH}(M-E_n)}, e^{S_{\rm rad}} \},
\label{eq:N_n}
\end{equation}
with $S_{\rm rad}$ the coarse-grained (thermal) entropy of the radiation.%
\footnote{We assume that the coarse-grained entropies of the three 
 components satisfy $S_{\rm hard} \ll S_{\rm soft} \approx S_{\rm BH}, 
 S_{\rm rad}$, which is expected to be valid throughout the (essentially) 
 whole history of black hole evolution.}
Here, we have performed the Schmidt decomposition in the space of soft-mode 
and radiation states for each $n$.  This expression makes it clear why 
the entanglement argument of Ref.~\cite{Almheiri:2012rt} does not apply 
here.  The entanglement responsible for unitarity has to do with the 
summation of the index $i_n$ shared between the soft-mode and radiation 
states (in fact, dominantly the vacuum index $i_0$), while the entanglement 
necessary for interior spacetime has to do with the index $n$, and these 
two can coexist.

The fact that Hawking emission occurs around the edge of the zone has 
an interesting implication for the nature of the horizon experienced by 
an infalling observer~\cite{Nomura:2018kia}.  Imagine that early Hawking 
radiation interacts with a detector, leading to different pointer states 
$\ket{d_I}$.  By separating these states from $\ket{\phi_a}$, the state 
in Eq.~(\ref{eq:sys-state}) can be written as
\begin{equation}
  \ket{\Psi(M)} = \sum_n \sum_{i_n = 1}^{{\cal N}(M-E_n)} 
    \sum_I \sum_{a_I} c_{n i_n I a_I} \ket{\{ n_\alpha \}} 
    \ket{\psi_{i_n}(M-E_n)} \ket{\phi_{a_I}} \ket{d_I}.
\label{eq:BH-occup-2}
\end{equation}
To discuss what the detector finding a particular outcome $I$ will 
experience later, we focus on the particular branch of the wavefunction
\begin{equation}
  \ket{\Psi_I(M)} = \frac{1}{\sqrt{z_I}} \sum_n 
    \sum_{i_n = 1}^{{\cal N}(M-E_n)} \sum_{a_I} c_{n i_n I a_I} 
    \ket{\{ n_\alpha \}} \ket{\psi_{i_n}(M-E_n)} \ket{\phi_{a_I}} \ket{d_I},
\label{eq:Psi_I}
\end{equation}
where $z_I = \sum_n \sum_{i_n = 1}^{{\cal N}(M-E_n)} \sum_{a_I} 
|c_{n i_n I a_I}|^2$ is the normalization factor.  Generically, this 
does not affect the physics of the black hole, since the structure 
of Eq.~(\ref{eq:Psi_I}) is the same as that of Eq.~(\ref{eq:sys-state}). 
However, if the detector is carefully set up, it may be fully 
correlated with a particular configuration $\{ n'_\alpha \}$ of 
the hard modes after the measurement:\ $c_{n i_n I a_I} \approx 0$ 
for $n \neq \{ n'_\alpha \}$.  This seems to mean that when the detector 
enters the horizon, it would hit a ``firewall'' because the hard modes 
lack the necessary entanglement.

However, since the detector can interact with semiclassical Hawking 
quanta only outside the zone, it can reach the stretched horizon only 
after time of order $4M l_{\rm P}^2 \ln(M l_{\rm P})$.  Therefore, if 
the equilibrium timescale between the hard and soft modes is of order
\begin{equation}
  t_{\rm eq} = 4Ml_{\rm P}^2 \ln(Ml_{\rm P})
\label{eq:t_eq}
\end{equation}
or shorter, then the state of the system (without the detector 
included) takes the form of Eq.~(\ref{eq:sys-state}) with generic 
$c_{n i_n a} = c_{n i_n I a_I}$ in the $n$ and $i_n$ spaces when the 
detector enters the stretched horizon.  This would imply that an operation 
acting only on early Hawking radiation---however complicated---cannot 
destroy the smoothness of the horizon one will see; a black hole 
self-repairs itself by the time an infaller reaches the horizon.

\section{Beyond Black Holes}
\label{sec:beyond}

In this section, we generalize the results of the previous section obtained 
for an evaporating black hole to other systems.  We first discuss how 
the physics of Rindler spacetime is obtained as a smooth limit of the 
black hole physics.  We then see that essentially all the ideas developed 
in the previous section can be applied consistently to de~Sitter spacetime. 
We finally consider asymptotically flat spacetime and discuss how some 
of the ideas developed here may be related to the semiclassical analysis 
of the asymptotic symmetry structure.

\subsection{Rindler limit}
\label{subsec:Rindler}

Rindler spacetime is obtained as a limit of Schwarzschild spacetime
\begin{equation}
  M \rightarrow \infty
\quad\mbox{and}\quad
  l_{\rm P}, l_{\rm s}: \mbox{ fixed}
\label{eq:Rind-limit}
\end{equation}
by focusing on the near horizon region, $r \rightarrow 
2Ml_{\rm P}^2$, such that the combinations
\begin{equation}
  \rho \equiv 2\sqrt{2Ml_{\rm P}^2(r-2Ml_{\rm P}^2)}
\qquad\mbox{and}\qquad
  \tau \equiv \frac{R}{4Ml_{\rm P}^2} t
\label{eq:rho-omega}
\end{equation}
are kept finite, where $R$ is a finite parameter with the dimension of 
length.  The resulting metric is
\begin{equation}
  ds^2 = -\frac{\rho^2}{R^2} d\tau^2 + d\rho^2 + \sum_i dX^i dX^i,
\label{eq:Rindler}
\end{equation}
where $X^i$ ($i=2,3$) parameterize the direction parallel to the horizon. 
Since the Rindler spacetime is obtained by taking the $M l_{\rm P} 
\rightarrow \infty$ limit of Schwarzschild spacetime, its entropy 
is infinite
\begin{equation}
  S_{\rm Rindler} = \infty,
\label{eq:S_Rindler}
\end{equation}
though the surface entropy density on the stretched horizon is still given 
by $1/4 l_{\rm P}^2$.

There is no direct analogue of Hawking emission in the Rindler limit, 
since the edge of the zone in the original Schwarzschild spacetime 
is at spatial infinity.%
\footnote{We implicitly imagine an IR cutoff $\rho_{\rm IR} \rightarrow 
 \infty$ so that $\rho_{\rm IR}/Ml_{\rm P}^2 < \infty$.}
There is, however, an analogue of black hole mining~\cite{Unruh:1982ic,%
Brown:2012un}, by which a physical probe at constant $r$ in the zone 
observes a thermal bath with temperature $T_{\rm loc}(r)$, which in the 
Rindler limit gives
\begin{equation}
  T_{\rm loc}(\rho) = \lim_{M \rightarrow \infty} 
    \frac{T_{\rm H}}{\sqrt{1-\frac{2Ml_{\rm P}^2}{r}}} 
  = \frac{1}{2\pi \rho}.
\label{eq:Rindler-T}
\end{equation}
This is the well-known Unruh effect~\cite{Unruh:1976db,Fulling:1972md,%
Davies:1974th}.  While mining allows us to extract information about a 
black hole vacuum, the Unruh effect is expected not to give any information 
about a microstate of the Rindler/Minkowski vacuum.  This is ensured 
by Eq.~(\ref{eq:S_Rindler}); since extracting information about 
a scrambled system requires accessibility to more than a half of 
its entropy~\cite{Page:1993wv,Hayden:2007cs}, no finite size detector 
can extract such information.

As in the case of a black hole, semiclassical theory in a Rindler wedge 
describes microscopic dynamics of only the hard modes, whose locally 
measured energies are sufficiently larger than $T_{\rm loc}(\rho)$. 
The other degrees of freedom, the soft modes, are described 
only statistically.  Denoting the states of the hard modes by 
$\ket{\{ n_\alpha \}}$, microstates of the vacuum can be written 
in the form of Eq.~(\ref{eq:BH-state}).  The construction of mirror 
operators and ``interior spacetime,'' i.e.\ the other side of the 
Rindler horizon, goes as in the black hole case (by first taking $M$ 
finite and then sending it to infinity).  An important difference, 
however, is that since there is no far mode, the mirror space is 
constructed purely out of the soft modes.

Another consequence of Eq.~(\ref{eq:S_Rindler}) is that the scrambling 
time~\cite{Hayden:2007cs,Sekino:2008he} of Rindler spacetime is infinite
\begin{equation}
  \tau_{\rm scr} \approx O\bigl(\rho\, \ln S_{\rm Rindler}\bigr) 
  \rightarrow \infty.
\label{eq:tau_scr}
\end{equation}
This has an operational meaning.  It implies that negative energy-entropy 
excitations, generated by backreaction of the Unruh effect and entangled 
with the detector, do not relax in any finite time.  If we reduce 
the acceleration characterizing the Rindler description, then this 
entanglement---information about the other side of the horizon---can be 
retrieved in the Rindler wedge.  While this can be viewed as an analogue 
of information retrieval from a black hole, Eq.~(\ref{eq:S_Rindler}) 
implies that the retrieved information is not scrambled.  This is 
consistent with the inertial frame description, which implies that 
the negative energy-entropy excitations, which can be viewed as particles 
emitted from the detector in an inertial frame~\cite{Unruh:1983ms}, 
are not thermalized when they reappear from the receding Rindler horizon.

\subsection{de~Sitter spacetime}
\label{subsec:dS}

A consistent microscopic description of de~Sitter spacetime is not yet known. 
There are, however, several proposals aiming toward it.  In particular, a 
description based on a holographic screen seems promising for describing 
cosmological de~Sitter spacetime~\cite{Nomura:2018kji,Nomura:2017fyh}, at 
least when the spacetime deviates---even slightly---from the pure de~Sitter 
vacuum (e.g.\ by the existence of another energy component beyond the 
cosmological constant).  Here instead of committing to a particular proposal, 
we assume the existence of a consistent description of (approximate) 
de~Sitter spacetime and study what the most straightforward extension 
of the black hole picture would imply for such a description.  For 
related descriptions of de~Sitter spacetime based on the static picture, 
see Refs.~\cite{Banks:2000fe,Banks:2010tj}.

The picture of a black hole in a distant frame is analogous to the static 
description of de~Sitter spacetime (with the radius inside out), whose metric 
is given by
\begin{equation}
  ds^2 = -(1 - H^2 r^2) dt^2 + \frac{1}{1 - H^2 r^2} dr^2 + r^2 d\Omega^2,
\label{eq:dS-metric}
\end{equation}
where $H$ is the Hubble parameter.  The entropy and the local temperature 
are given by~\cite{Gibbons:1977mu}
\begin{equation}
  S_{\rm GH}(H) = \frac{\pi}{H^2 l_{\rm P}^2},
\qquad
  T_{\rm loc}(r) = \frac{H}{2\pi} \frac{1}{\sqrt{1-H^2 r^2}}.
\label{eq:GH-S-T}
\end{equation}
Analogous to the black hole case, we separate modes of low energy quantum 
fields into hard, $\omega \gtrsim \Delta$, and soft, $\omega \lesssim 
\Delta$, modes, where
\begin{equation}
  \Delta \approx O(H)
\end{equation}
is taken to be sufficiently, e.g.\ $O(10)$, larger than $T_{\rm loc}(0) 
= H/2\pi$, and the frequency $\omega$ and $\Delta$ are both measured at 
$r = 0$.

The energy of the vacuum obtained by integrating the energy density (see 
Eqs.~(\ref{eq:soft-rho},~\ref{eq:E_soft,0})) is:
\begin{equation}
  E \sim N \int_0^{r_{\rm s}} T_{\rm loc}(r)^4 \sqrt{1-H^2 r^2} 
    \frac{4\pi r^2}{\sqrt{1-H^2 r^2}} dr 
  \sim \frac{1}{H l_{\rm P}^2},
\label{eq:dS-energy-prop}
\end{equation}
where $r_{\rm s}$ is the location of the stretched horizon (the string 
length away from the mathematical horizon $r = 1/H$)
\begin{equation}
  \frac{1}{H} - r_{\rm s} \sim H l_{\rm s}^2.
\label{eq:dS-stretched}
\end{equation}
Requiring that $\partial S_{\rm GH}(H)/\partial E = 1/T_{\rm loc}(0)$, the 
proportionality factor in Eq.~(\ref{eq:dS-energy-prop}) is determined to be 
unity:
\begin{equation}
  E = \frac{1}{H l_{\rm P}^2}.
\label{eq:dS-energy}
\end{equation}
While the relation of this energy---defined at $r = 0$ rather than 
asymptotic infinity---to more conventionally defined energies is not 
clear, it can be used to obtain a consistent semiclassical description 
of the system as we see below.

As in the case of a black hole, we regard the entropy $S_{\rm GH}(H)$ 
to represent the density of de~Sitter microstates, which correspond to 
different configurations of the soft modes.  A microstate representing 
the de~Sitter vacuum is then given by
\begin{equation}
  \ket{\Psi(H)} = \sum_n \sum_{i_n = 1}^{{\cal N}(n)} 
    c_{n i_n} \ket{\{ n_\alpha \}} \ket{\psi_{i_n}(n)},
\label{eq:dS-state}
\end{equation}
where $\ket{\{ n_\alpha \}}$ are orthonormal states of the hard modes 
with $\alpha$ collectively denoting the species, frequency, and 
angular-momentum quantum numbers of a mode, $E_n$ is the energy 
of the state $\ket{\{ n_\alpha \}}$ as measured at $r = 0$, and 
$\ket{\psi_{i_n}(n)}$ are orthonormal states of soft modes which have 
energy $E-E_n = 1/Hl_{\rm P}^2-E_n$.  Since $E \rightarrow E-E_n$ can 
be interpreted as $H \rightarrow H + E_n H^2 l_{\rm P}^2$,
\begin{equation}
  {\cal N}(n) = e^{S_{\rm GH}\left(H + E_n H^2 l_{\rm P}^2\right)} 
  \approx \exp\left(\frac{\pi}{H^2 l_{\rm P}^2} - \frac{2\pi E_n}{H}\right).
\label{eq:dS-cal-N}
\end{equation}
Assuming generic coefficients $c_{n i_n}$, we can trace out soft modes, 
which yields
\begin{equation}
  {\rm Tr}_{\rm soft} \ket{\Psi(H)} \bra{\Psi(H)}
  = \frac{1}{\sum_n e^{-\frac{E_n}{T_{\rm GH}}}} \sum_n 
    e^{-\frac{E_n}{T_{\rm GH}}} \ket{\{ n_\alpha \}} \bra{\{ n_\alpha \}}.
\label{eq:rho_dS}
\end{equation}
This is the thermal density matrix with the temperature
\begin{equation}
  T_{\rm GH} = T_{\rm loc}(0) = \frac{H}{2\pi}.
\label{eq:dS-T}
\end{equation}
It is remarkable that the understanding of the entropy and temperature 
in terms of soft modes is carried over without any modification from 
a black hole to de~Sitter spacetime.

States in which hard modes are excited are obtained by acting 
corresponding creation operators to a de~Sitter vacuum microstate. 
The annihilation and creation operators for hard modes take the 
form in Eqs.~(\ref{eq:ann},~\ref{eq:cre}):
\begin{align}
  b_\gamma &= \sum_n \sqrt{n_\gamma}\, 
    \ket{\{ n_\alpha - \delta_{\alpha\gamma} \}} \bra{\{ n_\alpha \}},
\label{eq:dS-ann}\\
  b_\gamma^\dagger &= \sum_n \sqrt{n_\gamma + 1}\, 
    \ket{\{ n_\alpha + \delta_{\alpha\gamma} \}} \bra{\{ n_\alpha \}}.
\label{eq:dS-cre}
\end{align}
It is these excitations of hard modes that we perceive as excitations 
over the de~Sitter vacuum at the semiclassical level.  The analysis 
in Section~\ref{subsec:paradox} for atypicality of excited states goes 
through in the de~Sitter case as well.

\subsubsection*{Outside the de~Sitter horizon}

\begin{figure}[t]
\begin{center}
  \includegraphics[height=6.5cm]{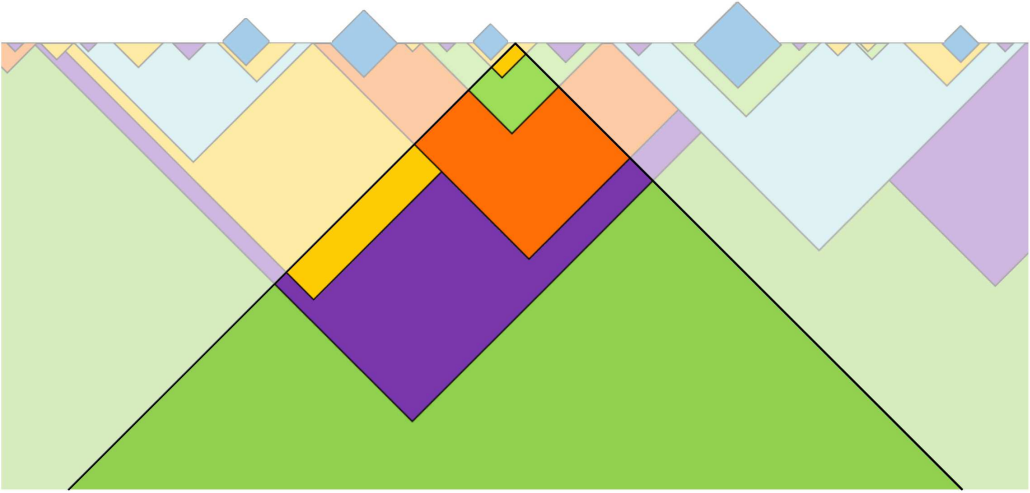}
\end{center}
\caption{A schematic depiction illustrating the idea that a single branch 
 of a quantum state describes only the spacetime region accessible by 
 a single observer.  The global spacetime of general relativity (the 
 entire region including those represented by lighter colors) arises 
 only as a ``pictorial depiction'' obtained by patching possible spacetime 
 histories represented by various branches.}
\label{fig:multiverse}
\end{figure}

In a realistic cosmological setup, de~Sitter spacetime appears approximately, 
and it is often the case that, in the standard general relativistic 
description, the region outside the horizon has much richer structure 
than the simple, plain de~Sitter space.  For example, if our universe 
began by a bubble nucleation in a parent universe, e.g.\ as one of 
infinitely many universes created in eternal inflation, then the Penrose 
diagram takes the form as in Fig.~\ref{fig:multiverse}.
The spacetime outside the de~Sitter horizon, or the holographic screen, 
has a complicated structure involving many other bubble universes, 
etc.  (Note that because our universe is not purely de~Sitter, e.g.\ 
by the existence of a small matter component or an early curvature 
dominated phase, the holographic screen at late times lies inside our 
bubble.)  On the other hand, if de~Sitter space with a fixed $H$ (specified 
within the precision allowed by the uncertainty principle) is described 
as a system with a finite number of degrees of freedom, then how can 
such a description be consistent with the possible existence of an 
innumerable variety of complicated spacetimes outside the horizon?

This problem was addressed in Ref.~\cite{Nomura:2011dt}, in which 
it was argued that the general relativistic, global spacetime as in 
Fig.~\ref{fig:multiverse} should (only) be interpreted as a pictorial 
depiction obtained by ``patching'' possible different semiclassical 
spacetime histories one can obtain in a quantum mechanical world, and 
that each of these histories describes only the spacetime region a single 
``observer'' (timelike curve) can access.  This leads to the following 
picture for the evolution of a quantum state.  As a standard scattering 
experiment converts an initial state with a specific particle configuration 
into a superposition of terms/branches with different particle 
configurations, a bubble nucleation---which is a quantum process---makes 
the state a superposition of branches having different spacetimes, 
e.g.\ with different bubbles created at different spacetime locations. 
Note that the state representing each branch may still have a finite 
coarse-grained entropy; in particular, for a branch in an approximate 
de~Sitter phase it is given by $S_{\rm GH}(H)$.  This therefore 
reconciles the finiteness of de~Sitter entropy with the ``existence'' 
of (infinitely) large spacetime outside the horizon in the general 
relativistic description.

The framework presented in this paper offers the possibility of making 
this picture more solid.  A specific question addressed is the following. 
While infinite spacetime outside the (approximate) de~Sitter horizon in 
a single branch may be an illusion, isn't it possible to access some part 
of it, e.g.\ when the system tunnels into a Minkowski vacuum or if slow-roll 
inflation ends with its potential energy converted into a different energy 
component by reheating?  How can a framework based on the static picture 
of Eq.~(\ref{eq:dS-metric}) describe such an ``information retrieval'' 
process?

Suppose that a state takes the form in Eq.~(\ref{eq:dS-state}) at $t = t_*$, 
possibly with hard modes excited by Eq.~(\ref{eq:dS-cre}).  As in the case 
of a black hole, we can erect an effective theory based on the state at 
$t = t_*$ by coarse-graining soft modes
\begin{equation}
  \sum_{i_n = 1}^{{\cal N}(n)} c_{n i_n} \ket{\psi_{i_n}(n)} 
  \longrightarrow\
  \frac{e^{-\frac{E_n}{2 T_{\rm GH}}}}
      {\sqrt{\sum_n e^{-\frac{E_n}{T_{\rm GH}}}}}\, \ketc{\{ n_\alpha \}}
\label{eq:dS-coarse}
\end{equation}
and introducing mirror operators
\begin{align}
  \tilde{b}_\gamma &= \sum_n \sqrt{n_\gamma}\, 
    \ketc{\{ n_\alpha - \delta_{\alpha\gamma} \}} 
    \brac{\{ n_\alpha \}},
\label{eq:dS-ann-m}\\
  \tilde{b}_\gamma^\dagger &= \sum_n \sqrt{n_\gamma + 1}\, 
    \ketc{\{ n_\alpha + \delta_{\alpha\gamma} \}} 
    \brac{\{ n_\alpha \}}.
\label{eq:dS-cre-m}
\end{align}
The vacuum state then becomes
\begin{equation}
  \ketc{\Psi(H)} 
  = \frac{1}{\sqrt{\sum_n e^{-\frac{E_n}{T_{\rm GH}}}}} 
    \sum_n e^{-\frac{E_n}{2 T_{\rm GH}}} 
    \ket{\{ n_\alpha \}} \ketc{\{ n_\alpha \}}.
\label{eq:dS-c-state}
\end{equation}
This allows us to interpret the mirror operators as representing 
semiclassical modes in the other hemisphere of spherical de~Sitter spacetime, 
with $\tilde{b}_\gamma^\dagger$ ($\tilde{b}_\gamma$) creating (annihilating) 
the mode that is the mirror image of the mode $\gamma$ in the original 
hemisphere with respect to the bifurcation surface.  The correspondence 
between the black hole and de~Sitter cases is as in Table~\ref{tab:corresp}. 
\begin{table}[t]
\begin{center}
\begin{tabular}{rc|c}
  & \mbox{Evaporating black hole} & \mbox{Cosmological de~Sitter space} \\
\hline
  \multirow{ 2}{*}{\mbox{microscopic level} $\Bigl\{$} 
  & \mbox{zone region} & \mbox{inside the horizon} \\
  & \mbox{far region}  & \mbox{-----} \\
\hline
  \multirow{ 2}{*}{\mbox{effective theory} $\Bigl\{$} 
  & \mbox{two-sided black hole} & \mbox{spherical de~Sitter space} \\
  & \mbox{the second exterior}  & \mbox{the other hemisphere}
\end{tabular}
\end{center}
\caption{Correspondence between an evaporating black hole and cosmological 
 de~Sitter space.}
\label{tab:corresp}
\end{table}
Note that for de~Sitter spacetime, there is no region corresponding to 
the region outside the zone of an evaporating black hole.

\begin{figure}[t]
\begin{center}
  \includegraphics[height=8.5cm]{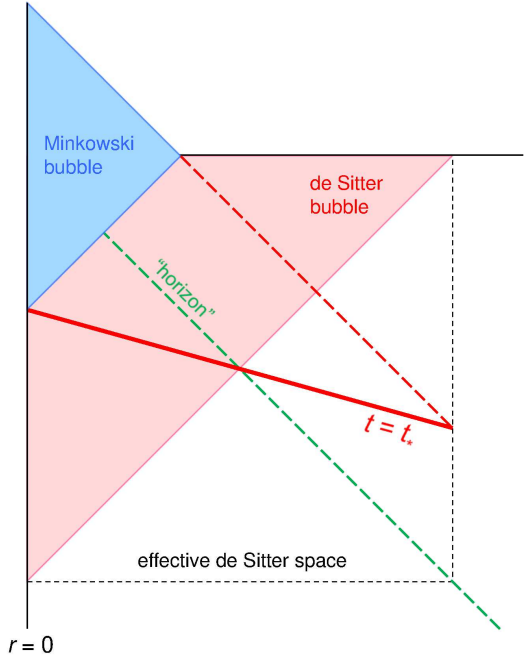}
\end{center}
\caption{For a branch having spacetime that appears approximately as a 
 static patch of a de~Sitter spacetime, an effective theory can be erected 
 by coarse-graining the soft modes at time $t = t_*$.  The resulting 
 theory contains a spatial section of an emergent full de~Sitter spacetime, 
 including the other side of the horizon.  This theory allows for describing 
 any future development of the branch.}
\label{fig:de-Sitter}
\end{figure}

As in the black hole case, the effective theory makes manifest the 
information about semiclassical physics encoded in collective excitations 
of the soft modes.  While the emergent spacetime region outside the horizon 
is finite, this is sufficient to describe any future development of the 
branch the system is in.  Suppose that at $t = t_*$ the coarse-grained 
state of the system is given by Eq.~(\ref{eq:dS-c-state}) with excitations 
on it.  This gives the state only on the $t = t_*$ hypersurface of the 
emergent de~Sitter space, depicted by the solid red line in the Penrose 
diagram in Fig.~\ref{fig:de-Sitter}.
Now, the observer associated with the branch (timelike curve at $r = 0$) 
can obtain the maximal amount of information about physics occurring 
outside the de~Sitter horizon (the dashed green line) if the system 
tunnels into a Minkowski vacuum right after $t = t_*$ as depicted in 
Fig.~\ref{fig:de-Sitter}.  We find that, even in this case, the knowledge 
about the state on the $t = t_*$ hypersurface is sufficient to fully 
describe the future signals the observer can receive.  In fact, since 
the argument relies only on causality, this implies that the effective 
theory can describe the future of the branch (as viewed from the observer) 
completely, even if its future history is more complicated, e.g.\ if 
the system evolves into a superposition of Minkowski bubbles created 
at different spacetime locations.

As the effective theory of the interior for an evaporating black hole, 
the effective theory discussed here is not unitary; for example, the 
future history of a particle that goes outside the causal domain of the 
$t = t_*$ hypersurface cannot be described.  As discussed above, however, 
complete physics the observer at $r = 0$ can access is described within 
the theory.   This can be done, for example, by relating the operators in 
Eqs.~(\ref{eq:dS-ann},~\ref{eq:dS-cre},~\ref{eq:dS-ann-m},~\ref{eq:dS-cre-m}) 
to the annihilation and creation operators for the flat slicing that 
describes the region inside the de~Sitter bubble:
\begin{align}
  a_\xi &= \sum_\gamma 
    \bigl( \alpha_{\xi\gamma} b_\gamma 
    + \beta_{\xi\gamma} b_\gamma^\dagger 
    + \zeta_{\xi\gamma} \tilde{b}_\gamma 
    + \eta_{\xi\gamma} \tilde{b}_\gamma^\dagger \bigr),
\label{eq:dS-a_xi}\\
  a_\xi^\dagger &= \sum_\gamma 
    \bigl( \beta_{\xi\gamma}^* b_\gamma 
    + \alpha_{\xi\gamma}^* b_\gamma^\dagger 
    + \eta_{\xi\gamma}^* \tilde{b}_\gamma 
    + \zeta_{\xi\gamma}^* \tilde{b}_\gamma^\dagger \bigr).
\label{eq:dS-a_xi-dag}
\end{align}
Here, $\xi$ is the label appropriate for the flat slicing, and 
$\alpha_{\xi\gamma}$, $\beta_{\xi\gamma}$, $\zeta_{\xi\gamma}$, and 
$\eta_{\xi\gamma}$ are the Bogoliubov coefficients calculable using 
the field theory method.

\subsection{Asymptotically flat spacetime --- relation to BMS}
\label{subsec:flat}

The analysis of Rindler and de~Sitter spacetimes described above has 
implications for the asymptotic structure of flat spacetime.  Let us 
take the flat space limit of de~Sitter spacetime:
\begin{equation}
  H \rightarrow 0
\quad\mbox{and}\quad
  l_{\rm P}, l_{\rm s}: \mbox{ fixed}
\label{eq:flat-limit}
\end{equation}
with $t$ and $r$ kept finite.  In this limit, the horizon is formally 
sent to infinity, which we may identify as spatial and null infinities 
of asymptotically flat spacetime.  The surface number density of the 
soft modes there is $1/4l_{\rm P}^2$.  In fact, the horizon appears 
locally as a Rindler horizon.

In the limit of Eq.~(\ref{eq:flat-limit}), the soft modes decouple from 
any experiment performed in a finite spacetime region, as indicated by 
the fact that the local temperature in Eq.~(\ref{eq:GH-S-T}) goes to zero 
for finite $r$
\begin{equation}
  T_{\rm loc}(r) \rightarrow 0.
\label{eq:flat-T}
\end{equation}
This implies that the infinite degeneracy of Minkowski vacuum, represented 
by different configurations of the soft modes, cannot be observed in 
any such experiment.  Note that the ``moduli space'' of this degeneracy 
is huge; it is formally given by the space of a unitary group
\begin{equation}
  {\cal M} \approx \left\lVert{ 
    U\biggl(\frac{\cal A}{4 l_{\rm P}^2}\biggr)} \right\rVert,
\label{eq:moduli-sp}
\end{equation}
where ${\cal A} = 4\pi/H^2 \rightarrow \infty$ is the area of the IR 
cutoff surface.  Given that each soft mode degrees of freedom can be 
arranged to lie at each Planck-sized region on the cutoff surface, 
we expect that these modes are related to the existence of the BMS 
group~\cite{Bondi:1962px,Sachs:1962wk,Strominger:2013jfa} and its 
possible generalization~\cite{Campiglia:2014yka} in perturbative 
quantum gravity.  This is consistent with the analysis in 
Refs.~\cite{Mirbabayi:2016axw,Bousso:2017dny} and our expectation 
that a physical process occurring in a finite spacetime region cannot 
determine the Minkowski vacuum microstate (otherwise, we would have 
to know a vast amount of information about the initial vacuum microstate 
to make predictions).

Note that this situation is different from that of a black hole in which 
the microscopic information stored in the soft modes can be extracted 
through a Hawking emission or mining process within a finite time.%
\footnote{A proposal relating black hole information to BMS soft hair 
 has been made in Ref.~\cite{Hawking:2016msc}.  Our framework is different 
 from this.  The black hole soft modes here contain a structure analogous 
 to but are not associated with the soft BMS charges at infinity. 
 In particular, the BMS symmetry or its extension at infinity does 
 not constrain the product of black hole evaporation.  See also 
 Refs.~\cite{Mirbabayi:2016axw,Bousso:2017dny} for a relevant discussion.}
The difference comes from the fact that a black hole is a finite system, 
interacting with a much larger system.  While holography entails that 
(a vast majority of) the information about a system is stored in its 
boundary region, the boundary of a black hole is located in the bulk 
of ambient space, reflecting interactions between the two systems. 
This allows for converting soft, microscopic information of the black 
hole into different configurations of the hard modes in the ambient 
space (or in the zone for a mining process), as was 
discussed in Section~\ref{sec:pic}.

\section{Discussion --- Observables in a Quantum World}
\label{sec:discuss}

We have seen that the formation of a horizon in quantum gravity is 
accompanied by the emergence of the soft modes arising from the large 
redshift.  While these modes cannot be discriminated temporarily by 
a semiclassical probe in the original reference frame, they play an 
essential role in describing the ``other side'' of the horizon, for 
example the interior of an evaporating black hole and the outside of 
a de~Sitter horizon.  These regions are described by effective theories 
in which specific operators involving soft modes, e.g.\ those in 
Eqs.~(\ref{eq:ann-m},~\ref{eq:cre-m},~\ref{eq:dS-ann-m},~\ref{eq:dS-cre-m}), 
play the role of annihilation and creation operators in the effective 
theories.  The construction in Sections~\ref{subsec:BH} and \ref{subsec:dS} 
guarantees that these operators always {\it exist}.  One might, however, 
still ask what selects them as ``good operators'' to describe the system, 
in particular the fate of an object entering the region behind a horizon. 
This has to do with the issue of the quantum-to-classical transition.

\subsubsection*{Quantum-to-classical transition: emergence of the Born rule}

Quantum mechanics is formulated to give probabilistic predictions for 
a measurement of a quantum system performed by a classical observer, 
whose existence is implicit in the Born rule.  Since the division between 
the quantum system and the surrounding classical world is arbitrary, we 
expect that the latter arises from more fundamental, intrinsically quantum 
mechanical substances.  While the precise mechanism for how a classical 
world emerges in quantum mechanics is unknown, it seems reasonable to expect 
that it has to do with amplification of information~\cite{q-Darwinism,%
q-Darwinism-2,Nomura:2011rb}, given that one of the most characteristic 
features of a classical system is the robustness of information.

A fundamental question about the emergence of a classical world is if 
it requires an infinitely large environment leading to truly irreversible 
decoherence, or if a sufficiently large environment holding (enormously) 
proliferated information is enough.%
\footnote{In cosmology, the former situation is realized in the scenarios 
 in Refs.~\cite{Nomura:2011dt,Bousso:2011up} while the scenario in 
 Ref.~\cite{Nomura:2012zb} requires the latter.}
If the former is true, then we might be able to declare that the only 
meaningful description of a system is the boundary one.  On the other 
hand, in the latter case, we expect that a description of the black hole 
interior must make sense, at least for a large enough black hole.  In 
the rest of the paper, we assume (as we have done so far) that the interior 
of a black hole can be consistently described in quantum mechanics.

In the standard treatment of quantum mechanics, it is customary to 
postulate that any Hermitian operator acting on the Hilbert space of 
the observed system is measurable.  This is reasonable if the surrounding 
system making observations has large resources, e.g.\ enough energy 
and controllability, so that the outcome of an observation associated 
with any such operator can be amplified and classicalized.  However, if 
the Hilbert space contains spacetime degrees of freedom, i.e.\ soft modes, 
then a process whose effects are fully contained in the corresponding 
spacetime region cannot measure everything about them, since there are 
not enough degrees of freedom within which outcomes are amplified.  For 
example, to measure all Hermitian operators acting on the soft modes 
of a black hole, we need to couple the black hole to a large external 
system, e.g.\ spacetime outside the zone, and use a process, e.g.\ 
Hawking emission, that allows for the amplification---and hence 
classicalization---of all possible outcomes.

The interior picture does not have such an external system,%
\footnote{The picture involves early Hawking radiation in the ambient 
 space, but this does not provide independent degrees of freedom that 
 can be used for amplification, as can be seen from the entanglement 
 structure in Eq.~(\ref{eq:ent-str}).}
and thus only a small portion of the operators are observable in the 
sense that they can be used in the Born rule by a classical observer 
who is a part of the system.  Suppose that the vacuum state at the 
coarse-grained level is given by
\begin{equation}
  \ketc{\Psi} 
  = \frac{1}{\sqrt{\sum_n e^{-\frac{E_n}{T}}}} 
    \sum_n e^{-\frac{E_n}{2 T} + i\varphi_n} 
    \ket{\{ n_\alpha \}} \ketc{\{ n_\alpha \}},
\label{eq:coarse-phase}
\end{equation}
where $\varphi_n = \varphi_{\{ n_\alpha \}}$ are phases.  The construction 
in Sections~\ref{subsec:BH} and \ref{subsec:dS} suggests that the 
observables correspond to Hermitian operators constructed out of 
the original creation and annihilation operators, $b_\gamma$ and 
$b_\gamma^\dagger$, as well as the mirror operators
\begin{align}
  \tilde{b}_\gamma &= \sum_n \sqrt{n_\gamma}\, 
    e^{i (\varphi'_{n_-} - \varphi'_n)} 
    \ketc{\{ n_\alpha - \delta_{\alpha\gamma} \}} 
    \brac{\{ n_\alpha \}},
\label{eq:ann-phase}\\
  \tilde{b}_\gamma^\dagger &= \sum_n \sqrt{n_\gamma + 1}\, 
    e^{i (\varphi'_{n_+} - \varphi'_n)} 
    \ketc{\{ n_\alpha + \delta_{\alpha\gamma} \}} 
    \brac{\{ n_\alpha \}},
\label{eq:cre-phase}
\end{align}
where $n_\pm \equiv \{ n_\alpha \pm \delta_{\alpha\gamma} \}$ and
\begin{equation}
  \varphi'_n = \varphi_n 
  \quad \forall n
\label{eq:phase-eq}
\end{equation}
(because the phases $\varphi_n$ can be absorbed by the redefinition of the 
coarse-grained states $\ketc{\{ n_\alpha \}} \rightarrow e^{-i \varphi_n} 
\ketc{\{ n_\alpha \}}$).  In fact, we can construct appropriately 
localized field operators out of these operators, and the generator 
of time evolution relating them can be given by Eq.~(\ref{eq:H_eff}). 
The state in Eq.~(\ref{eq:coarse-phase}) is then the ground state 
of this generator.

The fact that the operators in Eqs.~(\ref{eq:ann-phase},~\ref{eq:cre-phase}) 
provide the right building blocks is intuitive to understand.  As viewed 
from these operators---more precisely, operators related to them by 
a Bogoliubov transformation---the state in Eq.~(\ref{eq:coarse-phase}) 
represents a smooth spacetime, and we empirically know that information 
about the outcome of a measurement performed in the vicinity of such 
a state is appropriately classicalized when evolved by the generator 
of which Eq.~(\ref{eq:coarse-phase}) is the ground state.  On the other 
hand, if we choose the ``annihilation and creation operators'' of the 
form of Eqs.~(\ref{eq:ann-phase},~\ref{eq:cre-phase}) but violating 
Eq.~(\ref{eq:phase-eq}) in a generic manner, then the state in 
Eq.~(\ref{eq:coarse-phase}) appears as a ``firewall state'' with 
cutoff scale excitations, and Hermitian operators constructed out 
of them would not represent observables that can be used by a classical 
observer in the Born rule.

Given the role locality plays in the process of information 
amplification~\cite{q-Darwinism,q-Darwinism-2,Nomura:2011rb}, it seems 
reasonable to conjecture that the emergence of a classical world---in 
particular, a classical observer who can use the Born rule to predict 
the outcome of a measurement---requires that there exists a Hilbert space 
basis in which the Hamiltonian takes a local (nearest neighbor) form and 
the relevant states are sufficiently near the ground state so that the 
smooth spacetime picture is available.  This implies, for example, that 
for a state of the form in Eq.~(\ref{eq:sys-state}), the annihilation 
and creation operators appearing in observables must be taken to be
\begin{align}
  \tilde{b}_\gamma &= 
    \biggl( \sum_{n'} e^{-\frac{E_{n'}}{T_{\rm H}}} \biggr) 
    \sum_n \sqrt{n_\gamma}\,\, e^{-\frac{E_{n_-} + E_n}{2 T_{\rm H}}} 
\nonumber\\
  &\qquad \times 
    \sum_{i_{n_-} = 1}^{{\cal N}(M-E_{n_-})} 
    \sum_{i_n = 1}^{{\cal N}(M-E_n)} \sum_a \sum_b 
    c_{n_- i_{n_-} a}\, c^*_{n i_n b}\, 
    \ket{\psi_{i_{n_-}}(M-E_{n_-})} \ket{\phi_a} 
    \bra{\psi_{i_n}(M-E_n)} \bra{\phi_b},
\label{eq:ann-micro}\\
  \tilde{b}_\gamma^\dagger &= 
    \biggl( \sum_{n'} e^{-\frac{E_{n'}}{T_{\rm H}}} \biggr) 
    \sum_n \sqrt{n_\gamma + 1}\,\, e^{-\frac{E_{n_+} + E_n}{2 T_{\rm H}}} 
\nonumber\\
  &\qquad \times 
    \sum_{i_{n_+} = 1}^{{\cal N}(M-E_{n_+})} 
    \sum_{i_n = 1}^{{\cal N}(M-E_n)} \sum_a \sum_b 
    c_{n_+ i_{n_+} a}\, c^*_{n i_n b}\, 
    \ket{\psi_{i_{n_+}}(M-E_{n_+})} \ket{\phi_a} 
    \bra{\psi_{i_n}(M-E_n)} \bra{\phi_b},
\label{eq:cre-micro}
\end{align}
or those related in a simple way with these operators (e.g.\ by a 
Bogoliubov transformation or time evolution).  While we have not proven 
it, the conjecture seems plausible and would explain why the construction 
in Sections~\ref{subsec:BH} and \ref{subsec:dS} is adopted when describing 
the physics perceived by an object crossing the horizon.  It is the 
construction that makes manifest the observables to which the object 
can apply the Born rule.

The issue discussed here of selecting appropriate observables is irrelevant 
in asymptotically flat or AdS spacetimes because of the existence of an 
infinite amount of degrees of freedom at asymptotic infinity, to which 
the information can be amplified.  (In fact, one can view an asymptotically 
flat spacetime as a proxy of a sufficiently isolated system, with a physical 
observer ``modeled'' by the soft modes at infinity.)  In particular, 
the standard $S$-matrix paradigm does not---because it need not---address 
the issue.  However, cosmological spacetimes generally do not have such 
``infinitely powerful observers,'' reflecting the fact that the system 
is effectively finite.  Therefore, unless we resort to some infinite 
degrees of freedom somewhere, discussion about the quantum-to-classical 
transition---and hence the emergence of the Born rule---cannot be avoided. 
This indeed seems to be one of the most fundamental problems 
in understanding the world we live in.

\section*{Acknowledgments}

I am grateful to Pratik Rath and Arvin Shahbazi-Moghaddam for useful 
conversations.  I also thank the Yukawa Institute for Theoretical Physics 
at Kyoto University for holding the workshop YITP-T-19-03 ``Quantum 
Information and String Theory 2019,'' during which a part of this work 
was carried out.  This work was supported in part by the Department 
of Energy, Office of Science, Office of High Energy Physics under 
contract DE-AC02-05CH11231 and award DE-SC0019380, by the National 
Science Foundation under grant PHY-1521446, by MEXT KAKENHI Grant 
Number 15H05895, and by World Premier International Research Center 
Initiative (WPI Initiative), MEXT, Japan.

\end{document}